\documentclass[aps,reprint,twocolumn,showpacs,superscriptaddress,groupedaddress]{revtex4}  % for review and submission
\usepackage{graphicx}  % needed for figures
\usepackage{dcolumn}   % needed for some tables
\usepackage{bm}        % for math
\usepackage{amssymb}   % for math
\usepackage[sort&compress]{natbib}    % for bibliography

\begin{document}

% The following information is for internal review, please remove them for submission
%\leftline{Last edited \today}
%\leftline{Primary authors: Stephanie D. Song}
%\rightline{To be submitted to PRE Rapid Communication}
%\rightline{\em D\O\ INTERNAL DOCUMENT -- NOT FOR PUBLIC DISTRIBUTION}

% the following line is for submission, including submission to the arXiv!!
%\hspace{5.2in} \mbox{Fermilab-Pub-04/xxx-E}

\title{Extreme value analysis of gut microbial alterations in colorectal cancer}
\author{S.D. Song}
\affiliation{
Neuroscience Program, Wellesley College, 106 Central St., Wellesley, MA 02481\\
}
\author{P. Jeraldo}
\affiliation{
Microbiome Program, Center for Individualized Medicine, Mayo Clinic, 200 First St. SW, 
Rochester, MN 55905\\
}
\affiliation{
Department of Surgery, Mayo Clinic, 200 First St. SW, Rochester, MN 55905\\}
\author{J. Chen}
\affiliation{
Division of Biomedical Statistics and Informatics, Department of Health Sciences Research, 
Mayo Clinic, 200 First St. SW, Rochester, MN 55905\\
}
\author{N. Chia}
\email{Chia.Nicholas@mayo.edu}
\affiliation{
Microbiome Program, Center for Individualized Medicine, Mayo Clinic, 200 First St. SW, 
Rochester, MN 55905\\
}
\affiliation{
Department of Surgery, Mayo Clinic, 200 First St. SW, Rochester, MN 55905\\}

\date{\today}

\begin{abstract} % STEPHANIE
Gut microbes play a key role in colorectal carcinogenesis, yet reaching a consensus
on microbial signatures remains a challenge. This is in part due to a reliance on mean value
estimates. We present an extreme value analysis for overcoming these limitations. By characterizing 
a power law fit to the relative abundances of microbes, we capture the same microbial 
signatures as more complex meta-analyses. Importantly, we show that our method is robust to 
the variations inherent in microbial community profiling and point to future directions for 
developing sensitive, reliable analytical methods.
\end{abstract}

\maketitle

%%% Intro
Colorectal cancer (CRC) is the third most commonly diagnosed cancer in the United States 
\cite{Edwards2010}, resulting in an estimated 50,000 deaths annually \cite{Siegel2016}. 
Development of sporadic, i.e., non-hereditary, CRC is a complex process typically defined by 
the adenoma-carcinoma sequence, where there is first a transition from a normal colon 
epithelium to an adenomatous growth followed by a transition to a cancerous tumor 
\cite{Vogelstein1990}. Recently, evidence has been mounting that alterations in the gut 
microbiome--the approximately 100 trillion microbes residing in the gut--play a crucial role 
in this transition from normal epithelium to cancerous tumor 
\cite{Ahn2013,Zackular2013,Hale2017}.

Profiling the taxonomic composition of the gut microbiome has been made possible due to recent 
advances in 16S rRNA sequencing, a technique that quantitatively sequences hypervariable 
regions of the microbial rRNA present in a sample and assigns taxonomy accordingly 
\cite{Hayashi2002,Janda2007}. The microbial profiles of healthy subjects can then be compared to
the profiles of subjects with CRC. Specifically, one common approach is to identify taxa that are 
enriched in CRC subjects, then narrow down the list of microbes that potentially drive CRC progression. 
This may be sound in principle, but results are difficult to interpret in practice when
inconsistencies arise between different studies \cite{Zackular2013,Drewes2017}. 
For instance, oral pathogen \textit{Peptostreptococcus stomatis} is often considered a known 
associate of CRC, with several studies having corroborated these results 
\cite{Wang2012,Zeller2014,Feng2015,Nakatsu2015,Baxter2016,Yu2017,Shah2018}. Meanwhile, other 
studies show a very weak \cite{Ahn2013} or even no association at all 
\cite{Zackular2013,Castellarin2012,Kostic2012,Zackular2014,Flemer2017}.
At the same time, these studies often find a number of other possible signatures; thus, 
instead of clarifying the microbial drivers, these studies create additional confusion. 
Recent meta-analyses systematically analyze previously published 16S gene sequence data 
in an effort to identify consistent signatures. However, these results only confirm a limited number 
of species, and even highlight that the majority of results from individual studies
do not agree \cite{Shah2018,Drewes2017}.

One commonly proposed reason for the inconsistencies between studies is the existence of 
multiple mechanisms by which microbes can promote CRC 
\cite{Kostic2012,Zackular2013,Zackular2014,Drewes2017}. The 
result is that CRC drivers in one case may be uninvolved in another, with even the few 
confirmed CRC signatures displaying enrichment in, at most, a subset of cancerous stool 
or tumor biopsy samples 
\cite{Kostic2012,Castellarin2012,Zackular2014,Drewes2017,Flemer2017,Shah2018,Hale2018}. 
Typical analyses used to find these signatures rely on either the mean or a rank ordering 
of taxa abundances and therefore cannot reliably detect trends that only occur in a subset 
of samples. In other words, these measures only detect general and unidirectional 
shifts when, in fact, microbial effects are neither general nor linear. Further exacerbating 
these issues, mean value estimates often assume symmetrical Gaussian 
\cite{Bolker2009} or zero-inflated Poisson distributions \cite{Lambert1992,Xu2015}, which 
do not reflect real microbial distributions. The field is in crucial need of reliable analysis 
methods to parse out the signal from the noise. 

In this study, we make use of extremal distributions as a more sensitive and consistent means 
of identifying the major microbial signatures associated with CRC progression. In doing so, we 
show that the relative abundance distributions of putatively causative taxa follow a 
power law whose tail, i.e. extreme values, differs between normal, adenoma, and CRC samples. 
We use a permutation test with extreme value test statistics to quantify the differences 
in these asymmetrical distributions. We show our extremal analysis to be robust by 
replicating our findings in a separate, European cohort \cite{Zeller2014}, also corroborating 
the results of recent meta-analyses \cite{Drewes2017,Duvallet2017,Shah2018}. Employing a 
power law distribution to understand the role of potential microbial culprits in CRC 
progression will motivate, guide, and simplify future development of analytical methods for 
microbial data. 

%%% Discuss ubiquity power laws to motivate permutation test STEPHANIE
A power-law-like distribution of microbial relative abundances should not, in fact, be surprising. 
Observations of power laws are prevalent in nearly every field, and reflect
the universal importance of extreme events across fields ranging from economics to 
ecology, describing events from wealth distribution in the US to wildfire sizes \cite{Newman2005}. 
Even just within the field of physics, power law behaviors have received special attention; complex 
social networks \cite{Farkas2001}, avalanches \cite{Vives1994}, scientific citations \cite{Redner1998}, 
and football goals \cite{Greenhough2001} have all been characterized to follow power laws. 
While we cannot claim a power law distribution in our own work, understanding the characteristics 
of power laws can nonetheless guide the analysis of power-law-like data and reveal important insights. 
For example, one insight gained is that the extreme 
values comprising the heavy tail of power laws are often the most influential and informative to study 
\cite{Gabaix2003,Andriani2007}. A Gaussian approximation of a power law tail would underestimate the 
extreme values, thereby underestimating the mean. This insight deems approaches based on mean values 
less appropriate and calls for extreme value analysis (EVA).

% Results: Power Law
We examine the fecal microbiota of subjects who are healthy (n = 486), have adenomas (n = 
233), or have CRC (n = 17) by using 16S rRNA gene sequences from Hale et al. \cite{Hale2017}. 
The relative abundance of each taxon commonly cited in the field was determined for all samples at 
the genus level, the smallest taxonomic category able to be identified accurately. 
Average sequence read count was 6219, and samples with fewer than 2000 reads were removed as done 
in the original analysis by Hale et al. \cite{Hale2017}. Removing these samples protects
against the sensitivity to sequencing depth variations that relative abundances at the lower 
limit face.

\begin{figure} % Fig 1
\includegraphics[scale=0.58]{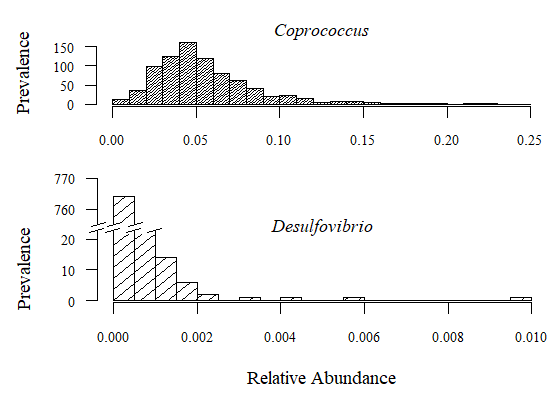}
\caption{\label{fig:extreme} Relative abundance distributions of typical high-abundance 
(\textit{Coprococcus}) and low-abundance (\textit{Desulfovibrio}) microbes. The former is more Gaussian, while the latter is heavily right-skewed.}
\end{figure}

While the relative abundances of high-abundance microbes show a distribution more similar to a 
symmetrical Gaussian distribution, low-abundance microbes display a highly right-skewed distribution 
with a heavy tail. Typical examples of genera displaying these two types of distributions are shown 
in Fig.~\ref{fig:extreme}.

\begin{figure*} % Fig 2 STEPHANIE
\includegraphics[scale=0.45]{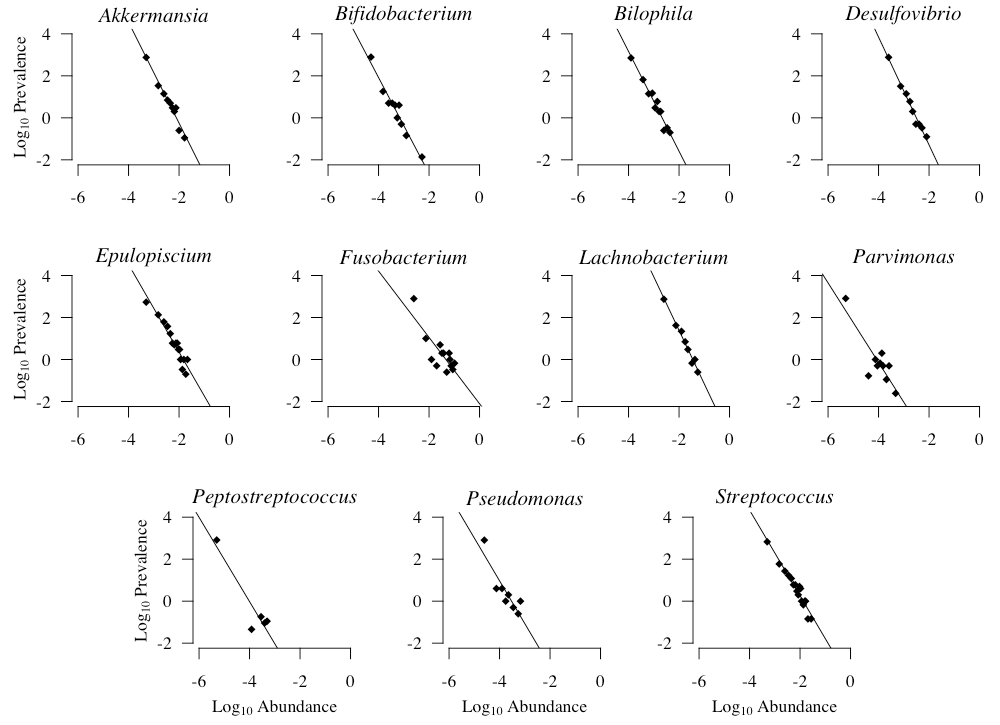}
\caption{\label{fig:loglog} Log-log transformations reveal that the 11 genera examined approximate
power law relative abundance distributions.}
\end{figure*}

As shown in Fig.~\ref{fig:loglog}, the right-skewed distributions do, in fact, approximate power 
laws. However, it should be noted that power laws are characterized by extreme behavior that extends 
to infinity, while relative abundances are bound by 1. In this study we only consider low-abundance 
genera with a mean relative abundance of less than 0.005. This left us with 11 out 
of the 18 total taxa characterized \footnote{Taxa characterized include \textit{Akkermansia, 
Bacteroides, Bilophila, Bifidobacterium, Clostridium, Coprobacillus, Coprococcus, Desulfovibrio, 
Epulopiscium, Fusobacterium, Lachnobacterium, Parvimonas, Peptostreptococcus, Prevotella, 
Pseudomonas, Ruminococcus, Streptococcus}, and \textit{Sutterella}.} (\textit{Akkermansia, 
Bifidobacterium, Bilophila, Desulfovibrio, Epulopiscium, Fusobacterium, Lachnobacterium, 
Parvimonas, Peptostreptococcus, Pseudomonas}, and \textit{Streptococcus}).

%%% Motivation for permutation test
Upon closer inspection, the length of the power law tail appears to differ according to disease status 
(data not shown). We therefore propose that the power law tails, i.e. extreme values, of CRC drivers 
are the values that differ between the normal, adenoma, and cancer stages of CRC progression. In order 
to assess this, we employ extreme value analysis to quantify these differences. Examining data from 
subjects at each stage of the adenoma-carcinoma sequence can then distinguish a microbe's role in the 
various stages of carcinogenesis \cite{Mori2018}. For example, a microbe that exhibits depletion of 
extreme values in normal samples but enrichment in adenoma samples would indicate a role in the early 
transition to cancer, whereas a microbe that exhibits only an enrichment in cancer samples would 
suggest a later effect.

\begin{figure} % Fig 3 STEPHANIE
\includegraphics[scale=0.45]{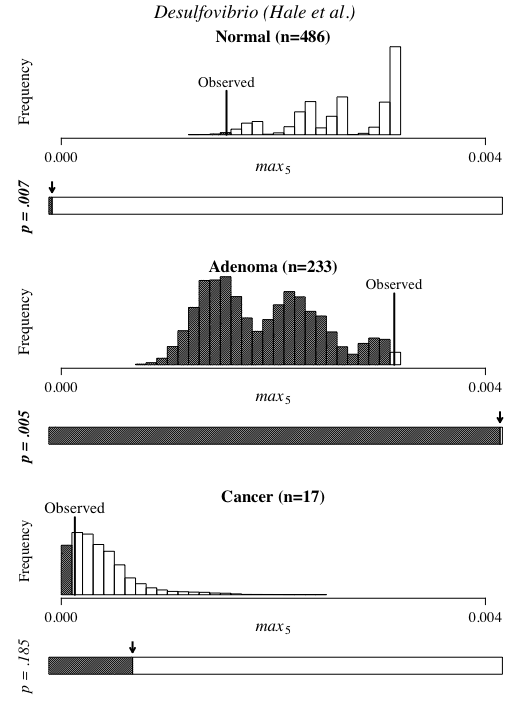}
\caption{\label{fig:permutation} Permutation distributions of the average of maximum 5 
relative abundances test statistic ($max_5$) for \textit{Desulfovibrio}. 10,000 random 
permutations were generated to produce expected distributions of the $max_5$ values 
in each group: normal, adenoma, and cancer. Lines indicate actually observed $max_5$ 
values. An observed value near the middle of the expected distribution would indicate an
expected result, while one near the ends would indicate significance. Consequently, observed 
$max_5$ is significantly lower than expected in normal and higher than expected in adenoma. 
In order to visualize the significance and direction of results, barcharts are shown 
below each histogram, with x-axis representing left-sided p-values ranging from 0 to 1. 
One-sided p-values (whichever is smaller) are also labeled on the left of the barcharts 
(bold p-values $< .05$). Therefore, a significantly lower than expected result is indicated 
by a small shaded region, while a significantly higher than expected result is indicated by
a large shaded region.}
\end{figure}

%%% method and results for permutation test
In order to focus our analysis on extreme rather than mean values, we perform a permutation 
test on an extreme value measure to determine the role of microbial taxa in CRC etiology. An 
advantage of using a permutation test is that it is non-parametric, hence we do not make 
faulty assumptions about distributional symmetry. We can also choose any extreme value measure 
to test. One such example is a simple maximum \cite{Fisher1928}; however, this statistic is 
overly dependent on a single data point, making it sensitive to random fluctuations and 
technical artifacts. Instead, we found averaging over the $x$ greatest values ($max_x$) to be 
a more reliable measure of extreme behavior. This statistic represents the maximum while eliminating 
noise, and it defines the "leading edge" of the abundance values. 

% STEPHANIE
While we utilize the previously mentioned insight about power law tails to focus our analysis on 
extreme values, importantly, our non-parametric approach does not rely on estimates of power law 
parameters, or even a rigorous claim that the relative abundances follow a strictly power law 
distribution.

We test this approach for different values of $x$ ($max_3$, $max_5$, and $max_7$) and find nearly 
identical results despite different sample sizes in each group (see Appendix A). These results indicate 
that these defined sizes of the leading edge are sufficient to produce reliable results, independent 
of sample size. We carry on by performing the permutation test using only the $max_5$ 
test statistic.  

By comparing the group $max_5$ values generated from the 10,000 random permutations of normal, 
adenoma, and cancer to the observed $max_5$ values, we are able to conclude whether the 
observed extreme values in each group are higher or lower than expected. Note that the calculated 
expectations for extreme values depend on sample size, as larger samples are more likely to 
include more extreme values than smaller samples. Correspondingly, cohorts with different sample 
sizes between groups lead to different expectations for extreme values in each group as well.

\begin{figure} % Fig 4
\includegraphics[scale=0.4]{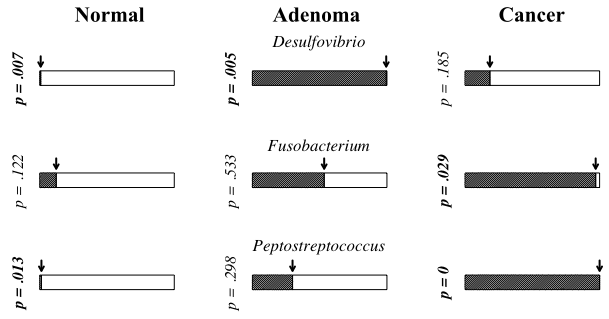}
\caption{\label{fig:haleresults} Taxa from Hale et al. with significantly different than expected 
extreme values at each stage of CRC progression. One-sided p-values are shown from the permutation 
test using the $max_5$ test statistic. Note that results fBold p-values $< .05$}
\end{figure}

When applied to \textit{Desulfovibrio}, we find that the $max_5$ statistic is lower than expected 
in the normal group and higher than expected in the adenoma group, indicating a role for this genus 
in the transition from normal to adenoma (Fig.~\ref{fig:permutation}). Applying these methods to 
the 10 other taxa studied, results are also found for \textit{Fusobacterium} and 
\textit{Peptostreptococcus}, both enriched in cancer (Fig.~\ref{fig:haleresults}).

\begin{figure*} % Figure 5
\includegraphics[scale=0.54]{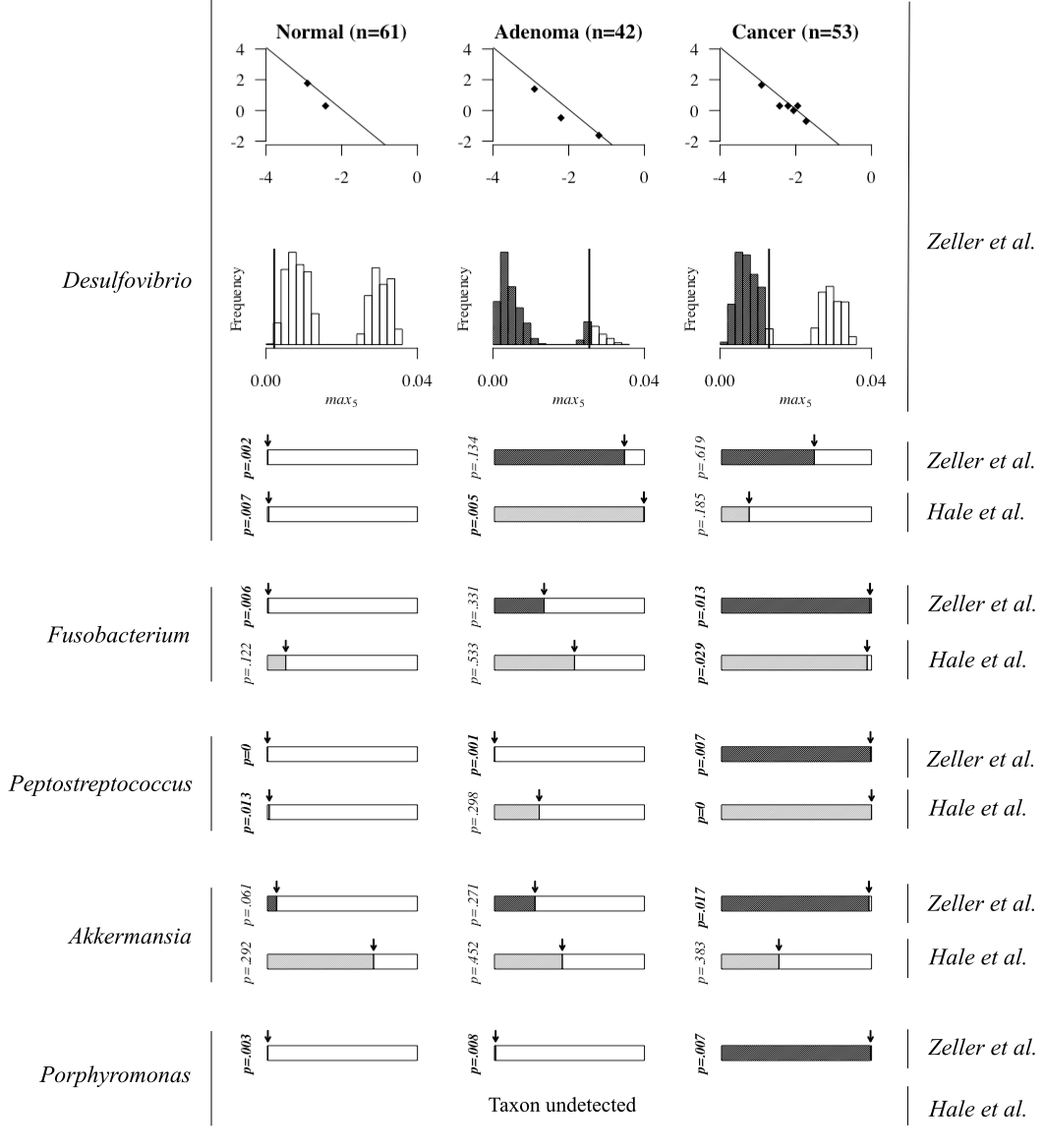}
\caption{\label{fig:combined} Replication of results in separate, European cohort (Zeller et al.) 
and comparison to previous Hale et al. results. Log-log transformations and permutation distributions 
for \textit{Desulfovibrio} using the $max_5$ statistic are shown as a representative example. 
One-sided p-values from the permutation test are visualized for both Zeller et al. and Hale et al. 
to compare. Note that results support consistent conclusions in both cohorts for 
\textit{Desulfovibrio}, \textit{Fusobacterium}, and \textit{Peptostreptococcus}. \textit{Porphyromonas} 
was undetected in Hale et al. Bold p-values $< .05$}
\end{figure*}

% Expected value test
As additional confirmation, we implement an exploratory parametric EVA method that explicitly 
applies the power law model to estimate an expected number of extreme values (see Appendix B 
for detailed methods). Briefly, we normalize a power law fit to the microbial abundances and 
estimate the number of expected extreme values above a defined critical relative abundance value, 
which we then compare to the observed number of extreme values. Our results match the permutation 
test results for \textit{Peptostreptococcus} ($p=0$ in the normal group, $p=.015$ in the adenoma 
group, $p=.009$ in the cancer group) and are non-significant but trending in the same direction for 
\textit{Desulfovibrio} ($p=.065$ in the normal group) and \textit{Fusobacterium} ($p=.141$ in 
the normal group). The uncertainty in the power law parameters lowers the detected significance, 
but overall these results strengthen our findings based on the permutation test. Improving the 
power law model and its corresponding PDF would further enhance the applicability of this approach. 

Finally, we examine if our results can be generalized to other studies. Inconsistencies between 
different studies arise due to many different factors including batch effects \cite{Leek2010,Chen2016} 
and sequencing biases from collection \cite{Sinha2016,Loftfield2016,Vogtmann2017}, primer design 
\cite{Jeraldo2011}, PCR conditions \cite{Gohl2016}, and lab-to-lab variation \cite{Sinha2015,Sinha2016}.
However, an advantage of our approach is that the overall distributional characteristics should remain 
relatively stable. We therefore expect our extreme value approach, which makes assessments based off 
the observed distribution, to be more robust to these variations. 

In order to test this, we replicate our results in a separate European cohort (61 normal, 42 adenoma, 
and 53 cancer) obtained from Zeller et al. \cite{Zeller2014} that used different collection methods, 
primer design, and sequencing platforms processed in a different lab \footnote{Data obtained from 
Zeller et al. was provided as relative abundances and average read length was not given. However, 
the original report indicates that samples with fewer than 1000 reads were removed 
\cite{Zeller2014}.}. The same 11 genera are tested as previously described using the permutation 
test. Indeed, results for all of the signatures found in Hale et al. (\textit{Desulfovibrio}, 
\textit{Fusobacterium}, and \textit{Peptostreptococcus}) are confirmed; \textit{Desulfovibrio} is 
found to be depleted in normal, and \textit{Fusobacterium} and \textit{Peptostreptococcus} enriched 
in cancer but depleted in normal (Fig.~\ref{fig:combined}).

It is worth addressing the two results that differ between the American and European
cohort. Specifically, the European cohort results show \textit{Akkermansia} and \textit{Porphyromonas} 
to have significant positive associations with cancer (Fig.~\ref{fig:combined}). 
However, these appear to be the result of differences in study
design. For \textit{Akkermansia}, cohort size of the cancer group was larger in the European cohort, 
making significance easier to assess in the adenoma to cancer transition. \textit{Porphyromonas} was 
completely undetected by Hale et al., making it methodologically impossible to assess computationally. 
In addition to study design-based limitations, it should also be noted that the p-values reported 
here are calculated without multiple hypothesis correction. However, the number of genera considered 
is relatively small, and our several replications of the results provide an additional level of 
stringency.

Overall, our results are very consistent, demonstrating the robustness of EVA. It is important
to note that our relatively simple permutation model performed on a single dataset captures many 
of the same signatures that otherwise have only been confirmed by more complex meta-analyses, 
most of which show agreement on the importance of \textit{Fusobacterium} and
\textit{Peptostreptococcus} \cite{Drewes2017,Duvallet2017,Shah2018}. The match in results is a 
testament to the equal, if not greater, sensitivity and reliability of EVA compared to other methods.
Moreover, EVA identified \textit{Desulfovibrio} and \textit{Akkermansia} to be significantly
associated with the transition to adenoma and cancer stage, respectively, a result that went 
undetected in the original analysis by Zeller et al. \cite{Zeller2014}.

Nonetheless, EVA has some limitations. By definition, EVA requires capturing rare 
events and thus performs best with a large number of samples. Fortunately, large-scale 
microbiome studies are becoming increasingly common \cite{HMP, Schmidt2018} as high-throughput 
sequencing becomes more accessible. In this context, EVA takes advantage of this progress while 
being less influenced by the variations inherent in microbial community profiling than mean 
value analyses.

% Discussion on how Power Law guides, simplifies, and motivates EVA
Future work may shift the paradigm of analytical methods for 16S rRNA sequencing data. 
Already, we have demonstrated the power of EVA to generate consistent results despite the 
multiple different mechanisms by which the microbiome drives CRC. Understanding that relative 
abundances of potential CRC drivers follow the ubiquitous power law distribution provides a 
guiding framework for developing future analytical tools. The characteristic lack of a finite 
mean exhibited by power laws explains the challenges of mean value analyses, motivating a 
shift towards EVA. Advancements in EVA, guided by this realization, will result in a clearer, 
simpler, and more accurate way to understand the role that key gut microbes play in the 
development and progression of CRC.

%%% Acknowledgements
We would like to extend our gratitude to Prof. Vanessa Hale for providing the data, Profs. 
Marc Tetel and Cassandra Pattanayak for continuing support and encouragement, and members 
of the Chia group and Prof. Marina Walther-Antonio for useful discussions. This work was 
supported an NIH grant (R01 CA 179243) as well as the Arnold and Mabel Beckman Foundation 
and Patterson Funds from the Neuroscience Program at Wellesley.

%%% Bibliography
\bibliographystyle{unsrtnat}
\bibliography{CRCBib}

\appendix
\section{Sensitivity of Extremal Averages}
\begin{figure}[b]
\includegraphics[scale=0.5]{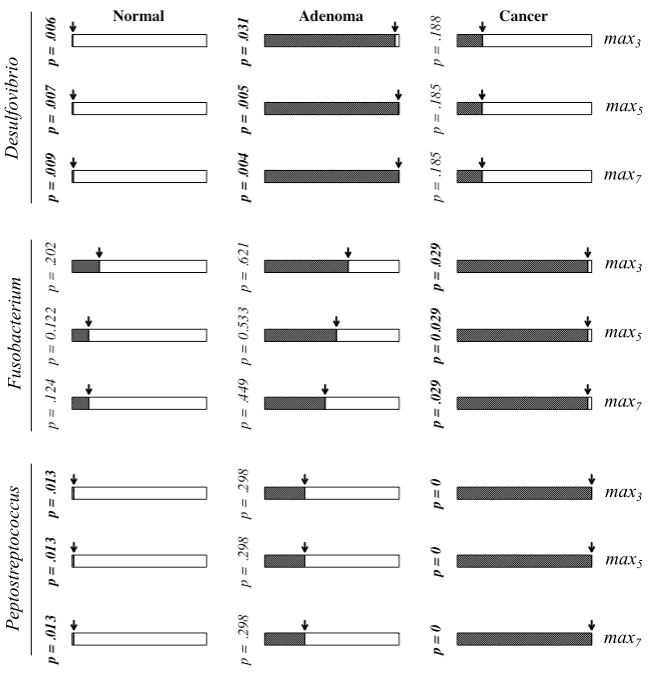}
\caption{\label{fig:appendix} Sensitivity analysis of permutation test statistic using 
average of maximum 3 ($max_3$), 5 ($max_5$), and 7 ($max_7$) values. Analyses were conducted
using data from \textit{Hale et al.} Bold p-values $< .05$}
\end{figure}

One method of measuring extreme values is to examine sample maxima; however, a single maximum value is 
often sensitive to random fluctuations and technical artifacts. Averaging a number of maximum values 
remedies this problem. Here, we considered using the average of 3 ($max_3$), 5 ($max_5$), and 7 
($max_7$) maximum values in order to derive the test statistic that is most suitable. All three 
measures performed nearly identically (Fig.~\ref{fig:appendix}). The index of dispersion was used as a 
quantitative measure of the differences between the test statistics,
\begin{equation}
D=\sigma^2/\mu 
\end{equation}
where $\sigma^2$ is the variance and $\mu$ is the mean of the three p-values for each group. Indeed, the resulting indices of dispersion were very low, ranging from $D$ = 0 to .014.

\section{Parametric EVA Methods}
The exploratory parametric EVA method first fits a power law to the microbial abundance data,
as in Fig.~\ref{fig:loglog}. $\beta_0$ and $\beta_1$ from a linear fit to the log-log transformed
data of bin interval $I$ are exponentiated to derive a crude power law fit:
\begin{eqnarray}
y=\beta_0+\beta_1x,\label{eqn:lr}\\
f(x)=10^{\beta_0}x^{\beta_1}\label{eqn:pl}
\end{eqnarray}
A cutoff point, or critical value, is then determined to define the extreme values. A conditional 
inference tree is used to find the critical value, and the average of 1000 bootstrap samples is taken.
We can now derive the total expected number of extreme values ($E$) from Eq.~(\ref{eqn:exp}), using the 
integral of the power law from the critical value to 1 (the maximum possible relative abundance):
\begin{eqnarray}
\int_{c}^{1}{f(x)dx}=\frac{10^{\beta_0}(1-c^{\beta_1+1})}{\beta_1+1},\label{eqn:exp}\\
E=\frac{\int_{c}^{1}{f(x)dx}}{I}\label{eqn:int}
\end{eqnarray}
The integral must be divided by bin interval $I$ because counts are discrete. The number of observed
extreme values in each group (normal, adenoma, and cancer) is then compared to expected using a 
binomial test, where $x=$ observed counts, $n=$ group sample size, and $p= E/$ total sample size. 

\end{document}